\documentclass{elsart4-1}

\usepackage{epsfig}

\usepackage{amssymb}

\usepackage[english,francais]{babel}

\newtheorem{e-proposition}[theorem]{Proposition}

\newtheorem{e-definition}[theorem]{Definition\rm}

\def\og{\leavevmode\raise.3ex\hbox{$\scriptscriptstyle\langle\!\langle$~}}
\def\fg{\leavevmode\raise.3ex\hbox{~$\!\scriptscriptstyle\,\rangle\!\rangle$}}

\begin{document}

\begin{frontmatter}


\selectlanguage{english}
\title{Dynamics of a trapped ultracold two-dimensional atomic gas}

\vspace{-2.6cm}

\selectlanguage{francais}
\title{Dynamique d'un gaz d'atomes ultra froid pi\'eg\'e \`a deux dimensions}


\selectlanguage{english}
\author[dgo]{David Gu\'ery-Odelin}
\ead{dgo@lkb.ens.fr}
\author[dgo]{Thierry Lahaye}
\ead{lahaye@lkb.ens.fr}

\address[dgo]{Laboratoire Kastler Brossel, Ecole normale sup\'erieure, 24 rue
Lhomond, F-75231 Paris Cedex 05, France}

\begin{abstract}
This article is devoted to the study of two-dimensional Bose gases
harmonically confined. We first summarize their equilibrium
properties. For such a gas above the critical temperature, we also
derive the frequencies and the damping of the collective
oscillations and we investigate its expansion after releasing of
the trap. The method is well suited to study the collisional
effects taking place in the system and in particular to discuss
the crossover between the hydrodynamic and the collisionless
regimes. We establish the link between the relaxation times
relevant for the damping of the collective oscillations and for
the time-of-flight expansion. We also evaluate the collision rate
and its relationship with the relaxation time.

\vskip 0.5\baselineskip

\selectlanguage{francais}
\keyword{Low-dimensional gas; Collective oscillations} ; Time of
flight}

\vskip 0.5\baselineskip \noindent{\small{\it Mots-cl\'es~:} Gaz en
basse dimension~; Oscillations collectives~;Temps de vol}

\end{abstract}
\end{frontmatter}

\selectlanguage{francais}
\section*{Version fran\c{c}aise abr\'eg\'ee}
Cet article est consacr\'e à l'\'etude des gaz de Bose \`a deux
dimensions en pr\'esence d'un confinement harmonique. Nous
abordons tout d'abord les propri\'et\'es d'\'equilibre de ce
syst\`eme. Nous d\'erivons ensuite, au-dessus de la temp\'erature
critique, l'expression des fr\'equences et de l'amortissement des
modes collectifs de basse \'energie et nous \'etudions avec le
m\^eme formalisme l'\'evolution du nuage d'atomes lorsque le
confinement est brutalement supprim\'e. La m\'ethode utilis\'ee
permet de d\'ecrire le gaz dans tous les r\'egimes collisionels,
du r\'egime sans collision au r\'egime hydrodynamique. Nous
\'etablissons le lien entre les temps de relaxation qui
d\'ecrivent les modes d'oscillation et l'expansion du nuage
apr\`es coupure du pi\`ege. Nous \'evaluons \'egalement
l'expression du taux de collision et sa relation avec le taux de
relaxation des modes de basse \'energie. \selectlanguage{english}
\section{Introduction}
\label{}

Quantum gases in reduced dimensionality are now experimentally
available. Two-dimensional Bose gases have been realized by
trapping atomic hydrogen at the surface of liquid helium
\cite{h2d,review}. Such a gas, confined into a box, undergoes at
sufficiently low temperature a superfluid transition known to be
of the Berezinskii-Kosterlitz-Thouless type
\cite{berez,Kost,svist}.

Recently, a few experiments with laser-cooled atoms have
approached the two-dimensional regime. The method used consists in
realizing very anisotropic confinement in  such a way that one
degree of freedom is frozen to zero motion oscillations. The
crossover to two-dimensions occurs when the thermal energy is
below the vibrational energy in the tightly confined direction.
Such a regime has been reached in standing-wave dipole traps
\cite{standing}. Alternatively, an elliptically focused laser beam
has been fed by a Bose-Einstein condensate leading to a
two-dimensional configuration \cite{gorlitz}. Evaporative cooling
of an atomic gas has also been performed at the crossover to two
dimensions in an optical surface trap \cite{hammes}. A
two-dimensional Bose-Einstein condensate has been achieved with
ultracold cesium atoms trapped in a gravito-optical surface trap
\cite{grimm03}. Notice that two-dimensional trapping by means of a
field-induced adiabatic potential has been recently proposed
\cite{zobay} and realized experimentally in the group of H. Perrin
and V. Lorent \cite{leshoucheshelene}. In all those systems the
two-dimensional confinement can be considered as harmonic.

The harmonic confinement introduces very different features from
the physics one would obtain in a box. As explained below, for an
ideal Bose gas, Bose-Einstein condensation
 occurs at finite temperature in the presence of a
harmonic potential in contrast to the homogeneous case
\cite{bagnato}.

The paper is devoted to the physics of two-dimensional Bose gases
harmonically trapped above the critical temperature. First, we
discuss the thermodynamic properties and the shift of the critical
temperature of an ideal gas at fixed number of particles due to
the finite size of the sample. We discuss briefly  the role of
interactions. Second, we study the dynamics of the gas through
low-lying collective modes and time of flight expansion by
including dissipative and mean-field effects.

\section{Thermodynamical properties}

\subsection{Ideal gas}

We consider $N$ bosons confined by a two-dimensional harmonic trap
of angular frequencies $\omega_x$ and $\omega_y$. By contrast with
a non-confined Bose gas, Bose-Einstein condensation is expected to
occur since the number of atoms in the excited states saturates to
an upper value $N'_{\rm max}$:

\begin{equation}
N'_{\rm max} =
\sum_{(n_x,n_y)\neq(0,0)}\frac{1}{\exp[\hbar(n_x\omega_x+n_y\omega_y)/(k_B
T)]-1}
\simeq\frac{\pi^2}{6}\frac{k_BT}{\hbar\omega_x}\frac{k_BT}{\hbar\omega_y}.
\label{eqnprime}
\end{equation}
The critical temperature for an ideal Bose-Einstein condensation
$T_c^0$ is readily obtained from this formula by setting $N'_{\rm
max}=N$:
\begin{equation}
k_BT_c^0=\frac{\hbar\bar{\omega}\sqrt{6 N}}{\pi},
\end{equation}
with $\bar{\omega}=(\omega_x\omega_y)^{1/2}$. Strictly speaking,
this value for the critical temperature is valid only in the
thermodynamic limit ($\bar{\omega}\rightarrow
0,\,N\rightarrow\infty$ with the product $\bar{\omega}^2 N$
keeping a constant value). We stress that due to the confining
potential,``true condensation" occurs in the sense that the first
excited states have a probability of occupation which tends to
zero in the thermodynamical limit.

\begin{figure}
\begin{center}
\epsfig{file=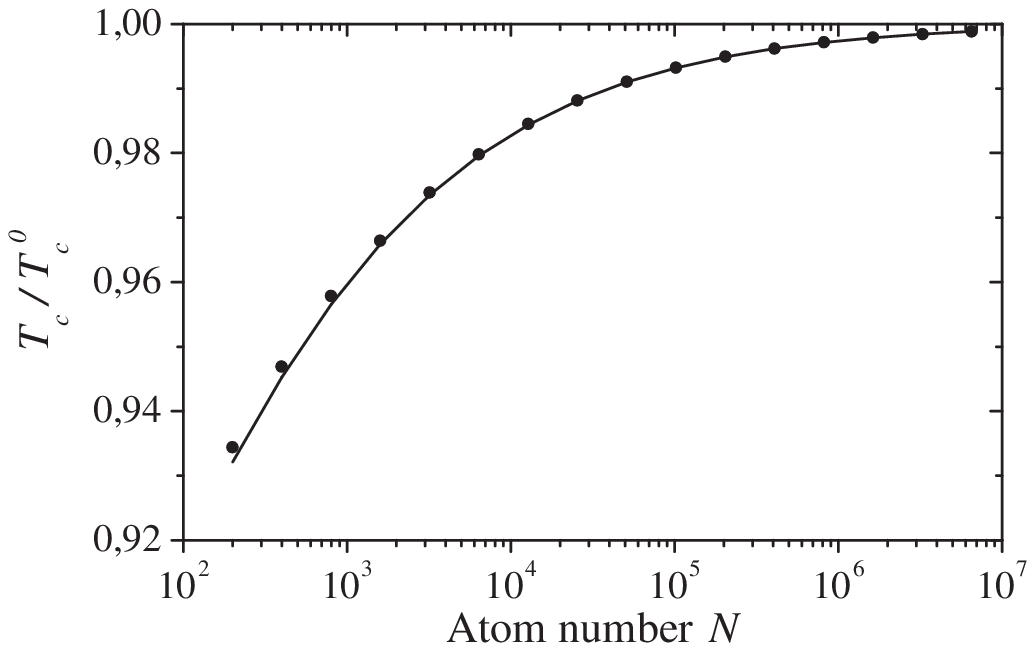, width=0.7\linewidth}
\begin{caption}
{Critical temperature $T_c$ (normalized to $T_c^0$) for a finite
number $N$ of atoms. The solid line is the analytical result given
by (\ref{eqlnn}), while the circles are given by numerically
solving for $T$ the equation $N'_{\rm max}(T)=N$. The agreement is
excellent even for low values of $N$.}
\end{caption}
\label{FigTofN}
\end{center}
\end{figure}

We find worthwhile  to examine the finite-size correction
\cite{finite} to the critical temperature as  real experiments are
more likely carried out with a small number of particles. The
scaling of the expected correction has been investigated in
\cite{mullin97}, we provide in the following a more quantitative
estimate. For a finite number $N$ of particles, the ``critical
temperature" $T_c$, still defined as the temperature for which
$N'_{\rm max} = N$, is shifted downwards. One can calculate
analytically, for an isotropic harmonic confinement, the leading
term correction for large but finite $N$ by evaluating the sum
(\ref{eqnprime}) more accurately (see appendix
\ref{appfinitesize}):
\begin{equation}
T_c=T_c^0 \left( 1-\frac{0.195\ln N-0.066}{\sqrt{N}} \right)\,.
\label{eqlnn}
\end{equation}
The prediction of Eq. (\ref{eqlnn}) is compared, on Fig.
\ref{FigTofN}, to the exact value of $T_c$ obtained by solving
numerically the equation $N'_{\rm max}(T)=N$. We notice that the
shift in $T_c$ is quite small by contrast with its
three-dimensional counterpart, even for $N$ as low as a few
hundred.

\subsection{Interactions}
\label{intf}

The interatomic collisions in a tightly confined Bose gas have
been studied in detail in \cite{petrov}. We summarize in this
section their main results. Similarly to the three-dimensional
scattering problem, the scattering properties in two dimensions
can be formulated by means of a single length $d^*$. This
characteristic length depends on the detailed shape of the
interatomic potential. For realistic momenta $q$ of particles in
ultracold gases we always have the inequality $qd^*\ll 1 $. The
expression for the two-dimensional differential cross section for
identical bosons reads: $d\sigma/d\theta(q)=|h(q)|^2/(4\pi q)$,
where $q=m({\bf v}_2-{\bf v}_1)/(2\hbar)$ is the relative
wavevector of the colliding particles and the dimensionless
function $h(q)$ depends logarithmically on $q$. The mean field
potential in the non degenerate regime is $U({\bf r})=2gn({\bf
r})$ with $n$ the density, $g=\hbar^2 h(\bar{q})/m$  and
$\bar{q}=(mk_BT)^{1/2}/\hbar$ the mean thermal wave vector.

The departure from the ideal gas behavior is measured through the
ratio between the mean field energy and the level spacing. For
weak interactions, namely $gn\ll \hbar\bar{\omega}$, interactions
can be taken into account perturbatively and the physics of the
ideal gas is valid. In the opposite limit $gn\gg
\hbar\bar{\omega}$, interactions brings about major differences.
One enters a regime of quasi-condensate \cite{shlyapq}. From
thermodynamics, one readily establishes the implicit equation for
the density:
\begin{equation}
n({\bf
r})=-\frac{1}{\lambda^2}\mbox{log}(1-\mbox{e}^{-\beta(U_{\rm trap
}({\bf r })+2gn({\bf r })-\mu)}), \label{dens}
\end{equation} where
$\lambda=h(2\pi mk_BT)^{-1/2}$ is the thermal de Broglie
wavelength, $\beta=1/k_BT$ and $U_{\rm trap }$ is the trapping
potential energy. Actually, in the non-perturbative interactions
regime, the local density approximation is valid. Repulsive
interactions tend to push the gas towards the external region,
thereby reducing the density of the gas. One expects that this
effect causes a natural decrease of the critical temperature. This
is indeed the case and in contrast with its three dimensional
counterpart, Eq. (\ref{dens}) is soluble at all temperatures
\cite{bhaduri}. To discriminate which state with or without
condensate the system will prefer, the authors of Ref. \cite{fern}
compare the free energy in both situations and predict that
low-energy phonons destabilize the two-dimensional condensate.
 The nature of the
phase in the degenerate regime remains somewhat an open question
\cite{mullin98,shlyapq,fern}. One may wonder if this state
involves superfluidity at sufficently low temperature.

\subsection{Expansion of the density}

In on-going experiments on Bose-Einstein condensates of alkali
atoms, the density profile is readily obtained by {\it in situ}
images of the cloud. We give in this section the corrections to
the density up to the second order due to the mean field
contribution. However, this expansion is valid not too close to
degeneracy. We expect that the repulsive mean field contributes to
the decrease of the density. We capture this effect perturbatively
by performing an expansion of the density profile with
$\xi=2g\beta\lambda^{-2}$ as the small parameter. $\xi$ is the
ratio between the mean field energy close to degeneracy  and the
temperature. By denoting $n^{(0)}_0$ the equilibrium density in
absence of mean field, we work out the expansion:
\begin{eqnarray}
&& n_0({\bf r})=n^{(0)}_0({\bf r})+\xi n^{(1)}_0({\bf
r})+\xi^2n^{(2)}_0({\bf r})+ ...\nonumber\\
&& n^{(0)}_0({\bf r})=-\lambda^{-2}\mbox{log}(1-\kappa)\nonumber\\
&& n^{(1)}_0({\bf
r})=-n^{(0)}_0\frac{\kappa}{1-\kappa}\nonumber\\
&& n^{(2)}_0({\bf r})=-\frac{n^{(1)}_0}{1-\kappa}\bigg(
\kappa-\frac{1}{2}\mbox{log}(1-\kappa)  \bigg),
\end{eqnarray}
with $\kappa=\kappa({\bf r})=\exp(-\beta (U_{\rm trap}({\bf
r})-\mu))$.

\section{Dynamics}
\subsection{Quantum Boltzmann equation}

The dynamics of the gas is described by a two-dimensional quantum
Boltzmann equation:
\begin{equation}
\frac{\partial f}{\partial t}+{\bf v_1}\cdot\mbox{\boldmath
$\nabla$}_{\bf r}f+\frac{{\bf F}}{m}\cdot\mbox{\boldmath
$\nabla$}_{\bf v_1}f=I_{\rm coll}[f] \label{Boltzmann}
\end{equation}
where $f$ is the phase space distribution function. The explicit
form for the collision term is
\begin{eqnarray}
I_{\rm coll}[f({\bf r},{\bf v_1})] &  \simeq  & \frac{m}{h}
\frac{|h(\bar{q})|^2}{4\pi^2}\int d\theta\; d^2{v}_2\;
\big[f({\bf r},{\bf v_{1^{\prime}}})f({\bf r},{\bf v_{2^{\prime}}})(1+f({\bf r},{\bf v_1}))(1+f({\bf r},{\bf v_2})) \nonumber \\
&&-f({\bf r},{\bf v_1})f({\bf r},{\bf v_2})(1+f({\bf r},{\bf
v_{1^{\prime}}}))(1+f({\bf r},{\bf v_{2^{\prime}}}))\big]
\nonumber
\end{eqnarray}
It accounts for elastic collisions between bosonic particles 1 and
2, with initial velocities ${\bf v_1}$ and ${\bf v_2}$, and final
velocities ${\bf v_{1^{\prime}}}$ and ${\bf v_{2^{\prime}}}$,
$\theta$ being the scattering angle in the center of mass frame.
In the next sections, we first derive the expression for the
collision rate with the interplay between statistics and mean
field. Then, we investigate the role of collisions (mean field and
relaxation) on the frequencies of the low-lying modes for a
two-dimensional harmonic trap and we derive the equations for
expansion after an abrupt switching off of the trap. The method
used to solve those latter problems relies on a scaling ansatz. It
is first introduced in the context of the collision-dominated
hydrodynamic regime, and then adapted to the quantum Boltzmann
equation \cite{pedri03}.

\subsection{Collision rate}
The collision rate plays a crucial role for the relaxation
dynamics, it permits to compare the mean free path to the size of
the cloud and to characterize the collisional regime of the sample
between collisionless and hydrodynamic. Well above the critical
temperature, the mean field and the bosonic statistics can be
neglected and the equilibrium distribution reads \cite{norm}:
\begin{equation}
f_0({\bf r},{\bf
v})=N.\frac{\hbar\omega_x}{k_BT}.\frac{\hbar\omega_y}{k_BT}e^{-H_0({\bf
r},{\bf v})/k_BT} \end{equation} with
\begin{equation}
 H_0({\bf r},{\bf
v})=\frac{1}{2}m{\bf
v}^2+\frac{1}{2}m\omega_x^2x^2+\frac{1}{2}m\omega_y^2y^2.
\end{equation}
We readily obtain the expression for the collision rate $\gamma_0$
in this limit by integrating the first term of the collision
integral of the classical Boltzmann equation for the equilibrium
function $f_0$ and dividing by the number of particles:
\begin{equation}
\gamma_0=\frac{m^3}{h^3}\frac{|h(\bar{q})|^2}{2\pi N}\int
d^2rd^2v_1\; d^2{v}_2\; f_0({\bf r},{\bf v}_1)f_0({\bf r},{\bf
v}_2)=\frac{N\hbar\omega_x\omega_y}{4\pi k_BT}|h(\bar{q})|^2.
\end{equation}

Closer to degeneracy but still above the critical temperature, one
expects that the result is modified by two opposite effects. On
the one hand, the mean field tends to decrease the density. On the
other hand, the statistics tends to shrink the cloud. To evaluate
the competition between those effects, we restrict ourselves in
the following to the case of an isotropic harmonic confinement for
sake of simplicity. We first expand the phase space distribution
function: $f({\bf r},{\bf v})=f_0({\bf r},{\bf v})[1+f_0({\bf
r},{\bf v})-2\beta gn_0({\bf r})]$. From the quantum Boltzmann
equation, the collision rate can be expanded to the first order
\begin{eqnarray}
\gamma & = & \frac{m^3}{h^3} \frac{|h(\bar{q})|^2}{4\pi^2 N}  \int
d\theta\; d^2{v}_2\;d^2{v}_1\;d^2{r}\; f_0({\bf r},{\bf
v_1})f_0({\bf r},{\bf v_2})\bigg[1+\nonumber \\ && f_0({\bf
r},{\bf v_1})+f_0({\bf r},{\bf v_2})+f_0({\bf r},{\bf
v_{1^{\prime}}})+f_0({\bf r},{\bf v_{1^{\prime}}})-4g\beta
n_0({\bf r})\bigg]. \label{gamm}
\end{eqnarray}
The signs in (\ref{gamm}) clearly show the enhancement of the
collision rate due to the statistics and its decrease due to the
repulsive mean field. After a lengthy but straightforward
calculation, we have worked out the explicit expression for the
expansion of the collision rate with respect to statistics and
mean field contribution:
$$
\gamma=\gamma_0\bigg[1+\frac{2\pi^2}{9}\bigg(\frac{T_c^0}{T}\bigg)^2\bigg(1-\frac{gm}{\pi\hbar^2}\bigg)\bigg].
$$
Actually, the collision rate cannot be measured directly. However,
all dissipative dynamics involve the collisions. We propose in the
following to extract information on the collisional regime through
the study of the collective oscillations or of the time-of-flight
expansion, which can be both investigated experimentally.

\subsection{Scaling ansatz for the hydrodynamic regime}
\label{hrs}

 In this section, we determine the frequencies of the
collective oscillations of an harmonically trapped two-dimensional
gas, and the equations for its time of flight when the confinement
is switched off. For this purpose, we transpose the hydrodynamic
approach developped in \cite{shlyap} for a three dimensional
trapped Bose gas to a two-dimensional one. The scaling ansatz for
the density reads: $n({\bf r},t)=n_0({\bf R}(t))/\Omega_0(t)$,
where $n_0$ is the equilibrium density, ${\bf R}=A(t).{\bf r}$
with ${\bf r}$ the coordinate and $A(t)$ a diagonal matrix with
time dependent scaling coefficient
$A=\mbox{diag}(1/a_x(t),1/a_y(t))$. The parameter $a_i$ gives the
dilation along the $i$th direction. One readily obtains
$\Omega_0(t)=(\mbox{det}(A))^{-1}$ from the normalization $\int
d^2r n({\bf r},t)=\int d^2r n_0({\bf r })$. Note that the scaling
solution makes sense only if $A$ has non vanishing coefficients.
The local hydrodynamic velocity field ${\bf v}^{\rm loc}$ is
obtained from the ansatz for the density through the equation of
continuity, which yields: ${\bf v}^{\rm loc}=-A^{-1}\dot{A}.{\bf
r}$, where $\dot{A}=dA/dt$. The equations for the coefficients of
the matrix $A$ are obtained from the Euler equation. For
collective oscillations, we find the following set of nonlinear
equations:
\begin{eqnarray}
\ddot{a}_x+\omega_x^2a_x-\frac{\omega_x^2}{a_x^2a_y}=0 \nonumber
\qquad\mbox{and}\qquad
\ddot{a}_y+\omega_y^2a_y-\frac{\omega_y^2}{a_y^2a_x}=0.
\label{hyds}
\end{eqnarray}
For a small deviation from equilibrium, the linearization of the
system leads to two hydrodynamic frequencies of oscillation:
$$
\omega^2_{\pm}=\frac{\omega_x^2}{2}\bigg[3+3\beta^2\pm
(9-14\beta^2+9\beta^4)^{1/2}\bigg]
$$
with $\beta=\omega_y/\omega_x$. For an isotropic trap
($\omega_x=\omega_y=\omega_0$), we find  $\sqrt{2}\omega_0$ for
the quadrupole mode and $2\omega_0$ for the monopole mode. For non
isotropic trap, the two modes correspond to a superposition of the
quadrupole and monopole modes. The equations for the time of
flight in two dimensions are readily obtained in the same way:
\begin{eqnarray}
\ddot{a}_x=\frac{\omega_x^2}{a_x^2a_y} \nonumber
\qquad\mbox{and}\qquad \ddot{a}_y=\frac{\omega_y^2}{a_y^2a_x}.
\end{eqnarray}
This set of equation shows that a high initial collision rate with
respect to the trap frequencies (implicitly assumed for the
validity of hydrodynamic equations) leads to an asymptotic
anisotropic expansion \cite{pedri03}.

\subsection{Scaling ansatz for the Quantum Boltzmann Equation}
In most cases the hydrodynamic formalism is not satisfactory for
describing a Bose gas. Indeed, a time of flight expansion is
accompanied by a dilution of the sample. As a result, the
collision rate decreases. Hydrodynamic equations are mostly valid
at the beginning of an expansion but certainly not after a long
time \cite{walraven}. We emphasize that the hydrodynamic regime is
quite difficult to reach experimentally \cite{mit,helium} since a
high collision rate means a high density for which the inelastic
collision rate is magnified. However by exploiting Feshbach
resonance, it has been possible recently to enter the hydrodynamic
regime for fermionic gases in a three dimensional trap
\cite{christophe}. In order to take into account the evolution of
the density one needs to solve at least approximately the
Boltzmann equation. This equation permits us to describe the
crossover between the collisionless and the hydrodynamic regime.

Following \cite{pedri03}, we make the following ansatz for the
distribution function: $ f({\bf r},{\bf v},t)=f_0({\bf R},{\bf
V})/\Omega(t)$ with $f_0$ the equilibrium distribution, ${\bf
V}=B.({\bf v}-{\bf v}^{\rm loc})$, $B$ being a diagonal matrix
$B=\mbox{diag}(1/b^{1/2}_x(t),1/b^{1/2}_y(t))$, and
$\Omega(t)=\mbox{det}^{-1}(AB)$. The parameter $b_i$ gives the
effective temperature in the $i$th direction and
 $f_0$ obeys the equilibrium Boltzmann equation:
\begin{equation}
\sum_j \bigg( V_j\frac{\partial f_0 }{\partial R_j
}-\omega_j^2R_j\frac{\partial f_0 }{\partial V_j
}-\frac{2g_0}{m}\frac{\partial n_0 }{\partial R_j }\frac{\partial
f_0 }{\partial V_j}\bigg)=0, \label{ebeq}
\end{equation}
where $n_0=(m^2/h^2)\int d^2V f_0$ is the equilibrium density and
$g_0$ the strength of the interaction for the equilibrium
temperature $T_0$. The scaling ansatz method does not provide an
exact solution of the Boltzmann equation. However, it is in
reasonable agreement with the numerical simulations based on
molecular dynamics \cite{pedri03,dgo99} and with experiments
\cite{orsay}. By substituting the scaling ansatz for the
distribution function in the Boltzmann equation and taking into
account the properties of the equilibrium distribution Eq.
(\ref{ebeq}) one obtains:
\begin{eqnarray}
&&\frac{1}{\Omega}\sum_i
\bigg[\bigg(\frac{b_i^{1/2}}{a_i}-\frac{g}{g_0}\frac{1}{a_ib_i^{1/2}}\frac{1}{\prod_ja_j}\bigg)V_i\frac{\partial
f_0 }{\partial
R_i}-\bigg(\frac{\dot{b}_i}{2b_i}+\frac{\dot{a}_i}{a_i}\bigg)V_i\frac{\partial
f_0 }{\partial V_i}-\nonumber \\
&&\bigg(\ddot{a}_i+\omega_i^2a_i-\frac{g}{g_0}\frac{\omega_i^2}{a_i}\frac{1}{\prod_ja_j}\bigg)\frac{R_i}{b^{1/2}_i}\frac{\partial
f_0 }{\partial V_i}\bigg]= I_{\rm
coll}+\frac{\dot{\Omega}}{\Omega^2}f_0. \label{eqb1}
\end{eqnarray}
This way, we restrict further the solutions. Such an approach
would fail for instance to take into account rotation in presence
of the mean field. Eq. (\ref{eqb1}) gives constraints on the
scaling coefficients $a_i$ and $b_i$. By denoting $\langle C
\rangle_0 = (m^2/h^2)\int C f_0({\bf R},{\bf V}) d^2Rd^2V/N$, we
readily derive the following set of equations by multiplying Eq.
(\ref{eqb1}) respectively by $C=R_iV_i$ and $C=V_i^2$ and by
performing the integration over the phase space:
\begin{eqnarray}
&&
\bigg(\frac{b_i}{a_i}-\frac{g}{g_0}\frac{1}{a_i}\frac{1}{\prod_ja_j}\bigg)\langle
V_i^2\rangle_0+
\bigg(\ddot{a}_i+\omega_i^2a_i-\frac{g}{g_0}\frac{\omega_i^2}{a_i}\frac{1}{\prod_ja_j}\bigg)
\langle R_i^2\rangle_0=0
\nonumber \\
&&\frac{\dot{b}_i}{b_i}+2\frac{\dot{a}_i}{a_i}=\frac{m^2}{h^2}\frac{\Omega}{N\langle
V_i^2\rangle_0}\int d^2Rd^2V V_i^2I_{\rm coll}. \label{syst}
\end{eqnarray}

 In order to
capture the physics of the collision integral, we treat this term
within  the relaxation approximation \cite{pedri03}. It allows us
to recast the l.h.s of the last equation of (\ref{syst}) in the
form:
$$
\frac{m^2}{h^2}\frac{\Omega}{N\langle V_i^2\rangle_0}\int d^2Rd^2V
V_i^2I_{\rm coll}=-
\frac{1}{b_i}\bigg(\frac{b_i-b_j}{2\tau}\bigg)\qquad\mbox{with}\qquad
i\neq j,
$$
where $\tau$ is the relaxation time which corresponds to the
average time between collisions. Using the properties of the
equilibrium distribution, we finally obtain the following closed
set of non linear equations for the scaling parameters:
\begin{eqnarray}
&&\ddot{a}_i+\omega_i^2a_i-\omega_i^2\frac{b_i}{a_i}+\omega_i^2\xi\bigg(
\frac{b_i}{a_i}-\frac{g}{g_0}\frac{1}{a_i}\frac{1}{\prod_ja_j}\bigg)=0
\nonumber \\
&&\dot{b}_i+2\frac{\dot{a}_i}{a_i}b_i=-\frac{1}{2\tau}(b_i-b_j)\qquad\mbox{with}\qquad
i\neq j,\label{syseq}
\end{eqnarray}
with $\xi=g_0\langle n_0\rangle_0/(m\omega_i^2\langle
R_i^2\rangle_0)$. One recovers the collisionless regime by taking
$\tau=\infty$. In this limit, we have a simple relation between
$b_i$ and $a_i$:
 $b_i=a_i^{-2}$. In the opposite limit (hydrodynamic regime), local equilibrium is
always ensured because of the high collision rate. As a
consequence, the contribution of the collision integral vanishes
because of local equilibrium and $b_x=b_y=1/(a_xa_y)$. As the mean
temperature is constant $g=g_0$, we recover in this limit the set
of equations (\ref{hyds}). In the next sections, we use this set
of equations to derive the frequencies of the collective
oscillations and we establish the equations for a time of flight
expansion.

\subsection{Collective oscillations}

Low lying collective oscillations are reproduced by the time
dependent dilatation parameters $a_x$ and $a_y$ of the trapped
cloud around equilibrium. As a consequence, the temperature
dependence of the interaction strength can be neglected $g=g_0$.
To evaluate the relaxation time $\tau_0$, at least at equilibrium
and in the high temperature limit, we perform a gaussian ansatz in
the collision integral as explained in details in \cite{dgo99} for
a three dimensional system. We find $\tau_0=2/\gamma_0$. Expanding
Eqs. (\ref{syseq}) around equilibrium ($a_i = b_i = 1$) we get a
linear closed set of equations which can be solved by searching
for solutions of the type $e^{i\omega t}$. The associated
determinant yields the dispersion law:
\begin{equation}
A[\omega]-\frac{i}{\tau_0}B[\omega]=0, \label{disp2}
\end{equation}
where $A[\omega]=\omega^2(\omega^2-\omega_{\rm cl
+}^2)(\omega^2-\omega_{\rm cl-}^2)$,
$B[\omega]=\omega(\omega^2-\omega_{\rm hd
+}^2)(\omega^2-\omega_{\rm hd -}^2)$ and (cl) and (hd) refer to
the collisionless and hydrodynamic regimes respectively. The
values for $\omega^2_{\rm cl \pm}$ and $\omega^2_{\rm hd \pm}$ are
given by
\begin{eqnarray}
\omega^2_{\rm cl \pm} & = & \frac{\omega_x^2}{2} \bigg[
(1+\beta^2)(4-\xi)\pm \{ (1+\beta^2)^2(4-\xi)^2-
32\beta^2(2-\xi)\}^{1/2}\bigg],
\nonumber \\
\omega^2_{\rm hd \pm} & = & \frac{\omega_x^2}{2} \bigg[
3(1+\beta^2)\pm \{9(1+\beta^4)-14\beta^2\}^{1/2}\bigg],
\end{eqnarray}with
 $\beta=\omega_y/\omega_x$.

The roots of $A$ that correspond to the collisionless regime in
presence of mean field coincide with the one derived by
\cite{guer}. The frequency of the monopole mode \cite{guer} for an
isotropic trap ($\omega_x=\omega_y=\omega_0$) is found to be
$2\omega_0$ whatever the collision regime. In two dimensions the
collisions (mean field and relaxation) do not affect the frequency
of this mode, this comes out from the fact that it is in the
kernel of the integral of collision \cite{dgo99}. The symmetries
of a two dimensional system for a contact interaction can explain
this surprising result \cite{rosch}. The hydrodynamic frequencies
do not depend explicitly on the mean field, it is also a specific
result of two dimensions low lying oscillations. As a consequence,
the effect of the mean field upon collective oscillations is
maximum in the crossover between collisionless and hydrodynamic
regime.

\subsection{Time of flight}

In the following we investigate the evolution of the cloud in a
time of flight. In this technique, the asymmetric trapping
potential is switched off and the evolution of the spatial density
is monitored. After a long time expansion, the  aspect ratio gives
information on the characteristics of the gas. Below the critical
temperature, the anisotropic expansion is the favored signature of
Bose-Einstein condensation. Actually, above the critical
temperature and in the collisionless regime one expects an
isotropic asymptotic expansion of the cloud reflecting the
isotropy of the initial velocity distribution. However,
anisotropic expansion also arises above the critical temperature,
when the mean free path is small compared to the size of the
cloud.

An analytic approach has been proposed in the full hydrodynamic
regime \cite{shlyap} for a three dimensional Bose gas and has been
adapted to a two dimensional gas in section \ref{hrs}.
Alternatively, the expansion of an interacting Bose gas initially
confined in a three dimensional trap above the critical
temperature has been investigated by means of Monte Carlo
simulations \cite{ARIMONDO}. Two recent papers
\cite{pedri03,walraven} propose a way to provide an analytical
interpolation between the two opposite collisionless and
hydrodynamic regimes. The authors of Ref. \cite{walraven} divide
the expansion in two stages: the first one is considered as fully
hydrodynamic and the second one as collisionless. The authors of
Ref. \cite{pedri03} use a scaling ansatz to solve approximately
the Boltzmann equation. We adapt in the following this latter
method to a two dimensional gas.

The same procedure as the one used previously for the collective
oscillations leads to the following set of equations:
\begin{eqnarray}
&&\ddot{a}_i-\omega_i^2\frac{b_i}{a_i}+\omega_i^2\xi\bigg(
\frac{b_i}{a_i}-\frac{g(\bar{b})}{g_0}\frac{1}{a_i}\frac{1}{\prod_ja_j}\bigg)=0
\nonumber \\
&&\dot{b}_i+2\frac{\dot{a}_i}{a_i}b_i=-\frac{1}{2\tau(a_i)}(b_i-b_j)\qquad\mbox{with}\qquad
i\neq j,\label{systof}
\end{eqnarray}
where $\bar{b}=(b_x+b_y)/2$. The confinement does not appear
explicitly as in (\ref{syseq}). The dependence with the parameters
$a_i$ and $b_i$ of the collision term through the interaction
strength $g$ and the time relaxation $\tau$ has to be taken into
account. As recalled in section \ref{intf}, $g$ depends
logarithmically on the mean temperature $T=T_0\bar{b}$. Since the
collision rate scales as the density, we deduce the scaling
dependence of the relaxation time: $\tau=\tau_0\prod_ia_i$. During
the expansion the scaling parameters $b_i$ decrease and as a
consequence the gas cools down, revealing the isentropy of the
expansion. By contrast with three-dimensional expansion, the mean
field contribution is slightly magnified during the expansion
through the dependence of $g$ upon the mean temperature. The
persistence of the hydrodynamic expansion that may happen
initially is more pronounced with respect to the three-dimensional
case because of the more favorable dependence of the relaxation
time on the scaling parameters.

\section{Conclusion}

In conclusion, our work shows that the coherent part of collisions
included through a mean field term can manifest itself in various
experimentally measurable ways already above the critical
temperature: by modifying the frequencies of low-lying modes, or
in competition with dissipative effects by producing an
anisotropic expansion for an initially anisotropic trapping.

\section*{ACKNOWLEDGMENTS}
We acknowledge fruitful discussions with M. Holzmann and useful
comments from H. Perrin.

\appendix

\section{Finite size effect on the critical temperature $T_c$}
\label{appfinitesize}

For $N$ bosons confined in a two-dimensional isotropic harmonic
oscillator of angular frequency $\omega$, the critical temperature
$T_c$ is the solution of the following equation
\begin{equation}
N = \sum_{n=1}^\infty\frac{n+1}{\exp[\hbar \omega n/(k_B
T_c)]-1}\,. \label{eqapp1}
\end{equation}
Since $T_c$ is close to $T_c^0=\hbar\omega\sqrt{6N}/(\pi k_B)$,
one has $\varepsilon\equiv\hbar\omega/(k_BT_c)\sim N^{-1/2}\ll 1$.

We therefore expand Eq. (\ref{eqapp1}) for $\varepsilon\ll 1$,
retaining only the lowest order terms, up to $1/\varepsilon$. The
sum can be replaced by an integral, provided we add a remainder
$\mathcal{R}(\varepsilon)$ to keep an exact expression:
\begin{equation}
N = \int_1^\infty \frac{x+1}{\exp(\varepsilon x)-1}{\rm d}x +
\mathcal{R}(\varepsilon)\,.
\end{equation}
The expansion of the integral is straightforward
\begin{equation}
\int_1^\infty \frac{x+1}{\exp(\varepsilon x)-1}{\rm d}x =
\frac{\pi^2}{6}\,\frac{1}{\varepsilon^2}-\frac{1}{\varepsilon}
-\frac{\ln\varepsilon}{\varepsilon}+\mathcal{O}(1)\,
\end{equation}
and the remainder can be evaluated with the Euler-McLaurin
asymptotic formula \cite{nr}
\begin{equation}
\mathcal{R}(\varepsilon)=-\sum_{k=1}^\infty\frac{B_k}{k!}f^{(k-1)}(1)\,\,\,\,\,
{\rm with}\,\,\,\,\,f(x)=\frac{x+1}{\exp(\varepsilon x)-1}
\end{equation}
where  $B_k$ is the $k$th Bernoulli number. One then expands the
derivatives of $f$ up to order $1/\varepsilon$, and one finds:
\begin{equation}
\mathcal{R}(\varepsilon)=\frac{S}{\varepsilon}+\mathcal{O}(1),
\end{equation}
where the coefficient $S$ is given by the following series:
\begin{equation}
S=1+1/12-1/120+1/252+\cdots
\end{equation}
which turns out to be a diverging series. Nevertheless, retaining
only a finite number of terms of this series actually gives a very
good approximation \cite{numrec} of the value of $S\simeq 1.079\pm
0.003$. Therefore,
\begin{equation}
N =
\frac{\pi^2}{6}\,\frac{1}{\varepsilon^2}+\frac{0.079}{\varepsilon}
-\frac{\ln\varepsilon}{\varepsilon}+\mathcal{O}(1)\,.
\end{equation}
Substituting $\varepsilon$ by $\sqrt{\pi^2/6N}(1-\delta
T_c/T_c^0)$ in the previous equation, and expanding up to first
order in $\delta T_c/T_c^0$ finally yields the shift of critical
temperature:
\begin{equation}
\frac{\delta T_c}{T_c^0}\simeq - \frac{0.195\ln N
-0.066}{\sqrt{N}}\,.
\end{equation}




\begin{thebibliography}{00}
\bibitem{h2d}
A.I. Safonov, S.A. Vasilyev, I.S. Yasnikov, I.I. Lukashevich, and
S. Jaakkola, Phys. Rev. Lett. {\bf 81}, 4545 (1998).

\bibitem{review}
J.T.M. Walraven, in {\it Fundamental Systems in Quantum Optics,
Proceedings of the Les Houches Summer School, Session LIII},
edited by J. Dalibard, J.M. Raimond, and J. Zinn-Justin (Elsevier
Science Publishers, Amsterdam, 1992).

\bibitem{berez}
V.L. Berezinski, Sov. Phys. JETP {\bf 32}, 493 (1971); V.L.
Berezinski, Sov. Phys. JETP {\bf 34}, 610 (1972).

\bibitem{Kost}
J.M. Kosterlitz and D.J. Thouless, J. Phys. C {\bf 6}, 1181
(1973); J.M. Kosterlitz,  J. Phys. C {\bf 7}, 1046 (1974).


\bibitem{svist}
N. Prokof'ev, O. Ruebenacker, and B. Svistunov, Phys. Rev. Lett.
{\bf 87}, 270402 (2001).



\bibitem{standing}
V. Vuleti\'c, C. Chin, A.J. Kerman, and S. Chu, Phys. Rev. Lett
{\bf 81}, 5768 (1998); I. Bouchoule, H. Perrin, A. Kuhn, M.
Morinaga, and C. Salomon, Phys. Rev. A {\bf 59}, 8(R) (1999); S.
Burger, F. S. Cataliotti, C. Fort, P. Maddaloni, F. Minardi and M.
Inguscio, Europhys. Lett. {\bf 57}, 1 (2002); I. Bouchoule, M.
Morinaga, C. Salomon, and D. Petrov
 Phys. Rev. A {\bf 65}, 033402 (2002).

\bibitem{gorlitz}
A. G\"orlitz, J.M. Vogels, A.E. Leanhardt, C. Raman, T.L.
Gustavson, J.R. Abo-Shaeer, A.P. Chikkatur, S. Gupta, S. Inouye,
T. Rosenband, and W. Ketterle, Phys. Rev. Lett. {\bf 87}, 130402
(2001).

\bibitem{hammes}
M. Hammes, D. Rychtarik, B. Engeser, H.-C. N\"agerl, and R. Grimm,
Phys. Rev. Lett. {\bf 90}, 173001 (2003).


\bibitem{grimm03}
D. Rychtarik, B. Engeser, H.-C. Nägerl and R. Grimm,
cond-mat/0309536.

\bibitem{zobay}
O. Zobay and B.M. Garraway, Phys. Rev. Lett. {\bf 86}, 1195
(2001); O. Zobay and B.M. Garraway cond-mat:0306198.

\bibitem{leshoucheshelene} to appear in the proceedings of the school
 {\it Quantum gases in low dimensions} at Les Houches 15-25 april 2003, Journal de Physique IV (EDP Sciences)
2003.

\bibitem{bagnato}
V. Bagnato and D. Kleppner, Phys. Rev. A {\bf 44}, 7439 (1991).

\bibitem{finite}
W. Ketterle and N.J. van Druten, Phys. Rev. A {\bf 54}, 656
(1996);

\bibitem{mullin97} W.J. Mullin, J. Low. Temp. Phys. {\bf 106}, 615
(1997).

\bibitem{petrov}
D.S. Petrov and G.V. Shlyapnikov, Phys. Rev. A {\bf 64}, 012706
(2001).

\bibitem{shlyapq} D.S. Petrov, M. Holzmann, and G.V. Shlyapnikov,
Phys. Rev. Lett. {\bf 84}, 2551 (2000).

\bibitem{bhaduri}
R.K. Bhaduri, S.M. Reimann, S. Viefers, A. Ghose Choudhury, J.
Phys. B {\bf 33}, 3895 (2000).

\bibitem{fern}
J.P. Fern\'andez and W.J. Mullin, J. Low. Temp. Phys. {\bf 128},
233 (2002).

\bibitem{mullin98} W.J. Mullin, J. Low. Temp. Phys. {\bf 110}, 167
(1998); W.J. Mullin, M. Holzmann and F. Lalo\"e, {\it ibid} {\bf
121}, 263 (2000); W.J. Mullin, M. Holzmann and F. Lalo\"e, {\it
ibid} {\bf 121}, 269 (2000).

\bibitem{norm}
We choose the following normalization $(m/h)^2\int d^2rd^2v f({\bf
r},{\bf v})=N$.

\bibitem{pedri03} P. Pedri, D. Gu\'ery-Odelin and S. Stringari, Phys. Rev. A.
{\bf 68}, 043608 (2003).

\bibitem{shlyap} Yu. Kagan, E.L. Surkov and G.V. Shlyapnikov,
Phys. Rev. A {\bf 55}, R18 (1997).

\bibitem{walraven}
I. Shvarchuck, Ch. Buggle, D.S. Petrov, M. Kemmann, W. von
Klitzing, G.V. Shlyapnikov, and J.T.M. Walraven,
arXiv:cond-mat/0308493.


\bibitem{mit}
D. M. Stamper-Kurn, H.-J. Miesner, S. Inouye, M. R. Andrews, W.
Ketterle, Phys. Rev. Lett. {\bf 81}, 500 (1998)

\bibitem{helium}
M. Leduc, J. L\'eonard, F. Pereira dos Santos, E. Jahier, S.
Schwartz and C. Cohen-Tannoudji, Acta Physica Polonica {\bf B} 33,
2213 (2002).

\bibitem{christophe}
T. Bourdel, J. Cubizolles, L. Khaykovich, K.M.F. Magalh\~aes,
S.J.J.M.F. Kokkelmans, G.V. Shlyapnikov, and C. Salomon, Phys.
Rev. Lett. {\bf 91} 020402 (2003).


\bibitem{dgo99}
 D. Gu\'ery-Odelin, F. Zambelli, J. Dalibard and
S. Stringari, Phys. Rev. A {\bf 60}, 4851 (1999).



\bibitem{orsay}
F. Gerbier, J.H. Thywissen, S. Richard, M. Hugbart, P. Bouyer and
A. Aspect, arXiv:cond-mat/0307188.


\bibitem{guer} D. Gu\'ery-Odelin, Phys. Rev. A {\bf 66}, 033613
(2002).

\bibitem{rosch} L.P. Pitaevskii and A. Rosch, Phys. Rev. A
{\bf 55}, R853 (1997).

\bibitem{ARIMONDO}
H. Wu and E. Arimondo, Europhys. Lett. {\bf 43}, 141 (1998).

\bibitem{nr}
W.H. Press, S.A. Teukolsky, W.T. Vetterling, B.P. Flannery, {\it
Numerical Recipes in C}, (2nd Edition, Cambridge University Press,
1992).

\bibitem{numrec} As explained in \cite{nr}, 4.2,
the error is smaller than the first neglected term of the
Euler-McLaurin series.

\end{thebibliography}
\end{document}